\documentclass[osajnl,twocolumn,showpacs,superscriptaddress,10pt,floatfix]{revtex4-1} 
\usepackage{amsmath,amssymb,graphicx}
\usepackage{color}
\begin{document}

\title{Universal Axial Fluctuations in Optical Tweezers}

\author{Marco Ribezzi-Crivellari}
\affiliation{Departament de F\'isica Fonamental, Universitat de Barcelona, Diagonal 645, Barcelona}

\author{Anna Alemany}
\affiliation{Departament de F\'isica Fonamental, Universitat de Barcelona, Diagonal 645, Barcelona}

\author{Felix Ritort}\email{Corresponding author: fritort@gmail.com}
\affiliation{Departament de F\'isica Fonamental, Universitat de Barcelona, Diagonal 645, Barcelona}
\affiliation{Ciber-BBN de Bioingener\'ia, Biomateriales y Nanomedicina, Instituto de Salud Carlos III, Madrid}

\begin{abstract}
Optical tweezers allow the measurement of fluctuations at the nano-scale, in particular fluctuations in the end-to-end distance in single molecules. 
Fluctuation spectra can yield valuable information, but they can easily be contaminated by instrumental effects. 
We identify axial fluctuations, i.e. fluctuations of the trapped beads in the direction of light propagation, as one of these instrumental 
effects. Remarkably, axial fluctuations occur on a characteristic timescale similar to that of conformational (folding) transitions, which may lead to misinterpretation of the experimental results. We show that a precise measurement of the effect of force on both axial and conformational fluctuations is crucial to disentangle them. Our results on axial fluctuations are captured by a simple and general formula valid for all optical tweezers setups and provide experimentalists with a general strategy to distinguish axial fluctuations from conformational transitions.
\end{abstract}


\maketitle 
A central application of optical tweezers (OTs) is found in the field of single-molecule manipulation, where they are used to exert force on single biopolymers (DNA, RNA and proteins) and measure extension changes in real time. The spatial and temporal resolution of OTs have increased considerably in the last decades. State-of-the-art OTs have now Angstrom resolution on a large bandwidth \cite{Nature.abbondanzieri.2005}. 
These important technological advances make it possible to resolve hitherto unobservable small and fast conformational transitions 
in tethered molecules. OTs are well suited to measure time-dependent fluctuations at the micro and the nano-scale, not just on single molecules \cite{Prl.meiners.2000} but also on single cells \cite{mizuno2009high} and in microrheology experiments \cite{Prl.crocker.2000}.

In this paper we will consider noise measurements on single molecules with OTs. We identify three
main noise components which shape the spectra of such measurements: elastic fluctuations, i.e. fluctuations in the end-to-end
distance of the molecule due to the flexibility of biopolymers (Fig. 1a), conformational fluctuations, such as folding/unfolding transitions (Fig. 1b), and axial fluctuations, i.e. fluctuations in the direction of light 
propagation (Fig. 1c). While the first two components convey information about the tethered molecule, the last one is instrumental and affects, although to a different extent, all OT setups.
Axial fluctuations are often significant because both the trapping effect and the restoring force due to the molecular
tether are weak in the axial direction. Like any noise source, axial fluctuations contribute to the variance of bead position fluctuations and thus limit the resolution of position measurements. Moreover, due to the low stiffness in the axial direction, the decay rate of axial fluctuations can be close to that of conformational fluctuations and distinguishing between these two noise sources may be difficult. In the following we shall establish a general criterion to assess whether a low frequency noise component is due to axial or to conformational fluctuations.
\begin{table*}\label{tabulla} 
  \caption{Results of the fits in Fig.~3a}
\centering
  \begin{tabular}{l|c|c|c|c|c|c|c|c}
    \hline
    \hline
   Tether, Setup & $\gamma_z^{-1} k_z$\, (s$^{-1}$)\, & $\gamma_z^{-1} \ell^{-1}$\, (pN s)$^{-1}$\, & $\ell$\, ($\mu$m)\, & $r_B,x_m$ ($\mu$m) & $\gamma^{-1}_z$\,($\mu$m pN$^{-1}$s$^{-1}$) \,  & $k_z$\, (pN/$\mu$m)\, \\
    \hline
    \hline
    Barnase, ST & $160\pm11$ & 19$\pm$1 & 2.0\, & $1.7,0.3$  & 39$\pm$2 & 5$\pm$1\\
    Short Handles, ST & $170\pm20$ & 17$\pm$1 & 1.7 & $1.7,\leq 10^{-2}$ & 29$\pm$1 & 6$\pm$1 \\
    Beads only, ST & 150$\pm$10 & - & - & - & 32$\pm$2 & 6$\pm$1 \\
    \hline
    24 kbp, DT & 45$\pm$3 & 2.7$\pm$0.4 & 12\footnote{Note that in the DT case $\alpha=2$.} & 2,8 & 16$\pm$3 & 2.7$\pm$0.4   \\
    3 kbp, DT & 31$\pm$10 & 7.1$\pm$0.9 & 5\footnotemark[1]& 2,1 & 17$\pm$3 & 1.7$\pm$0.6 \\
    Short handles, DT & 50$\pm$10 & 7$\pm$1 & 4\footnotemark[1] & $2,\leq 10^{-2}$ & 14$\pm$2 & 3$\pm$1 \\
    Beads Only, DT & 30$\pm$10 & - & - & - & 15$\pm$2 & 2.2$\pm$0.3 \\
    \hline
    \hline
    \end{tabular}
\end{table*}

\begin{figure}[htp]
\centerline{\includegraphics[width=.65\columnwidth]{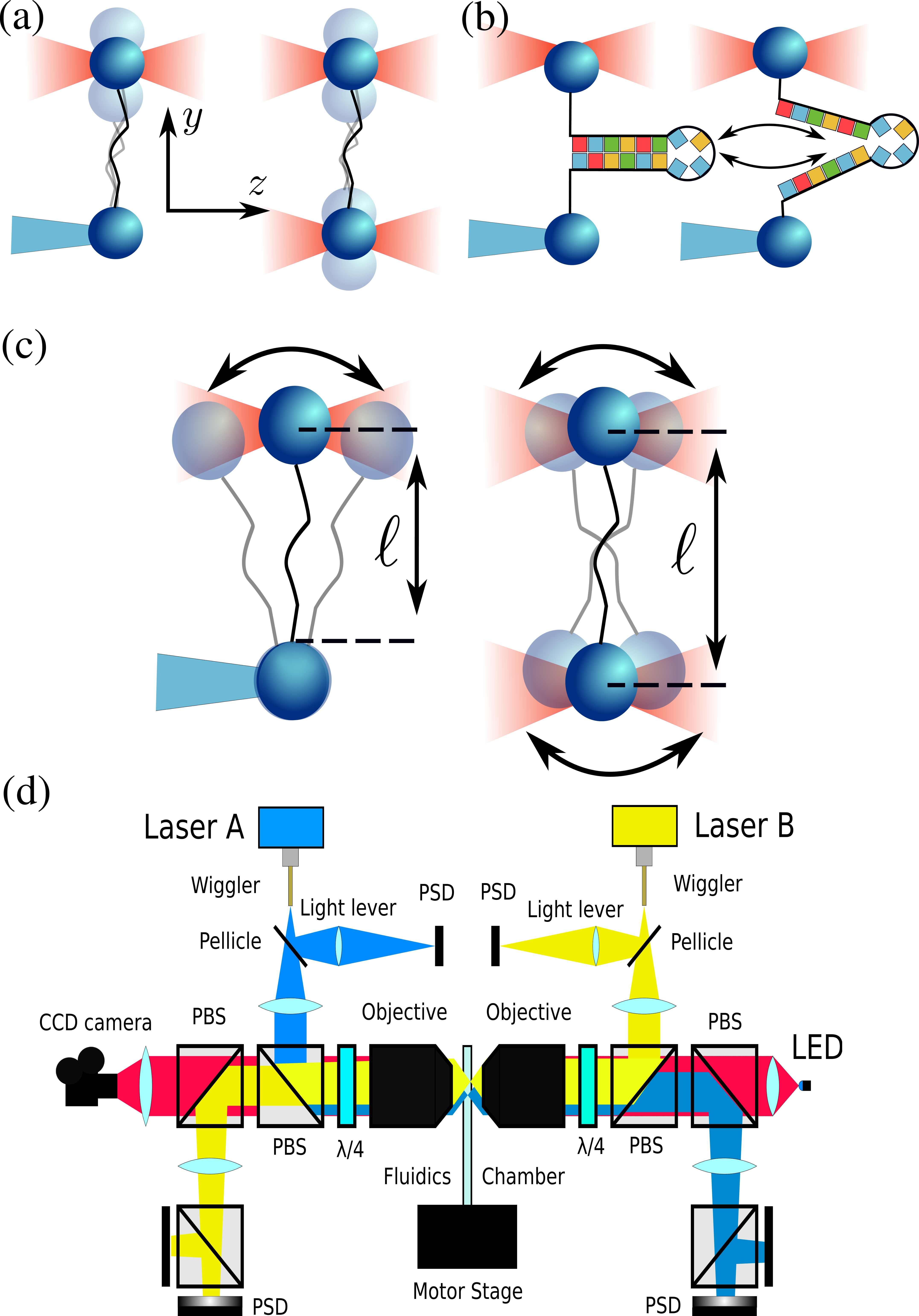}}
\caption{{\bf Fluctuation modes in OT experiments.}  Elastic (a), conformational (b) 
and axial (c) fluctuations are depicted. Elastic and axial fluctuations are shown in both the single and the dual-trap setup. d) Scheme of the counter-propagating dual-trap optical tweezers setup used in the experiments.  }
\end{figure}
We will discuss fluctuation spectra obtained in an OT setup which can operate both in the single \cite{PNAS.huguet.2010} and dual trap \cite{ribezzi2013counter} configurations and directly measures force via linear momentum conservation \cite{MEnzy.smith.2003} (Fig.~1d). Fluctuation spectra in OTs are most often used to measure the trap stiffness \cite{berg2004power,van2010power} In Fig.~2a we show the spectrum of position fluctuations of an optically trapped micron-sized bead recorded at high bandwidth ($\sim$20kHz).  As expected for Brownian motion in an harmonic well such spectrum can be fitted to a Lorentzian curve:
\begin{equation}
S(\nu)=\frac{k_BT}{2\pi^2\gamma}\frac{1}{\nu^2+\nu_c^2},
\end{equation}
where $\nu_c$ is the corner frequency, $k_B$ is the Boltzmann constant, $T$ is the absolute temperature and $\gamma$ the viscous friction affecting the trapped bead. The corner frequency is proportional to the decay rate $\omega_c$, $\omega_c=2\pi\nu_c$, and the decay rate is the inverse of the typical timescale of fluctuations, $\omega_c=\tau^{-1}_c$. The decay rate is determined by the ratio of the transverse trap stiffness, $k$, to $\gamma$: $\omega_c=k/\gamma$. In practice important corrections in the spectrum, Eq. (1), may appear at high frequencies due to detector transparency \cite{berg2004power}. 
From the spectrum one can measure the stiffness and thus convert position measurements to force measurements or vice versa. In OT single-molecule experiments the trapped bead is used to exert force on a biomolecule in either the single or dual trap configurations (Fig.~1a-c) and the dynamics of the bead becomes strongly coupled to that of the molecule. In this situation the fluctuation spectrum does not necessarily satisfy Eq. (1). As an example we consider experiments on DNA hairpins with short handles \cite{BiophysJ.forns.2011}. When a DNA hairpin is held under force, it can perform thermally induced transitions between the folded and the unfolded states (Fig.~1b). These transitions can be observed because they induce a change in molecular extension $x_m$ (Fig.~2b, upper trace) which is transmitted to the bead. 
In an ideal setup axial fluctuations would be absent. In this situation the fluctuation spectrum would take two superimposed contributions: conformational fluctuations between the folded and the unfolded states (corresponding to the big extension jumps shown in the upper trace of Fig. 2b) and elastic fluctuations  within one state, due to the flexibility of the tethered molecule.
Elastic fluctuations are Gaussian in the overwhelming majority of cases and have a Lorentzian spectrum. They yield information about the tether stiffness, in the same way as, in the absence of tethers, position fluctuations of a trapped bead can be used to measure the trap stiffness (Fig.~2a) \cite{Prl.meiners.2000,BiophysJ.forns.2011,ribezzi2012force}. Moreover they set the limit to the temporal resolution of the setup: the complex made by the beads and the tethered molecule can be thought of as a low-pass filter whose corner frequency equals that of elastic fluctuations. In fact, conformational fluctuations can be observed whenever they happen on a timescale that is larger than that of elastic fluctuations and have sufficient amplitude. For two state folders conformational fluctuations can be modeled by a telegraphic (dichotomic) noise and do also show a Lorentzian spectrum.
Two extreme cases are shown in Fig.~2b: in the upper (center) trace we show the behavior of a long $\sim$ 20bp (short $\sim$ 6bp) two-state DNA hairpin. 
In the first case conformational fluctuations are easily identified (telegraphic/dichotomic signal), while in the second case they are almost completely masked by elastic fluctuations. The corresponding spectra, shown in Fig.~2c (circles (squares) for the upper (center) trace) reflect the different nature of the two traces. Both spectra can be fit to a double Lorentzian (continuous lines) with the slow (lower corner frequency) component corresponding to conformational transitions and the fast component corresponding to elastic fluctuations. In the spectra obtained for the long hairpin (circles in Fig.~2c) the two Lorentzian shoulders are clearly distinguishable. However, for the shorter hairpin (squares in Fig.~2c), the corner frequency of elastic and conformational fluctuations come closer and
the amplitude of conformational fluctuations is greatly reduced.
In real experiments a third noise source is unavoidable: axial fluctuations of the trapped object. This kind of fluctuations has been well
characterized in magnetic tweezers studies, where they are used for force calibration \cite{gosse2002magnetic}.
As for the other components, also axial fluctuations have a Lorentzian spectrum. In Fig.~2b (lower trace) we show the signal obtained from a DNA hairpin at high force, where conformational transitions do not take place. Even in the absence of conformational fluctuations the spectrum has a double Lorentzian shape (Fig.~2c, diamonds), just as for the other spectra in Fig.~2c. However, in this case, the low-frequency Lorentzian shoulder (arrow) is due to axial fluctuations. 
\begin{figure}[htp]
\centerline{\includegraphics[width=.75\columnwidth]{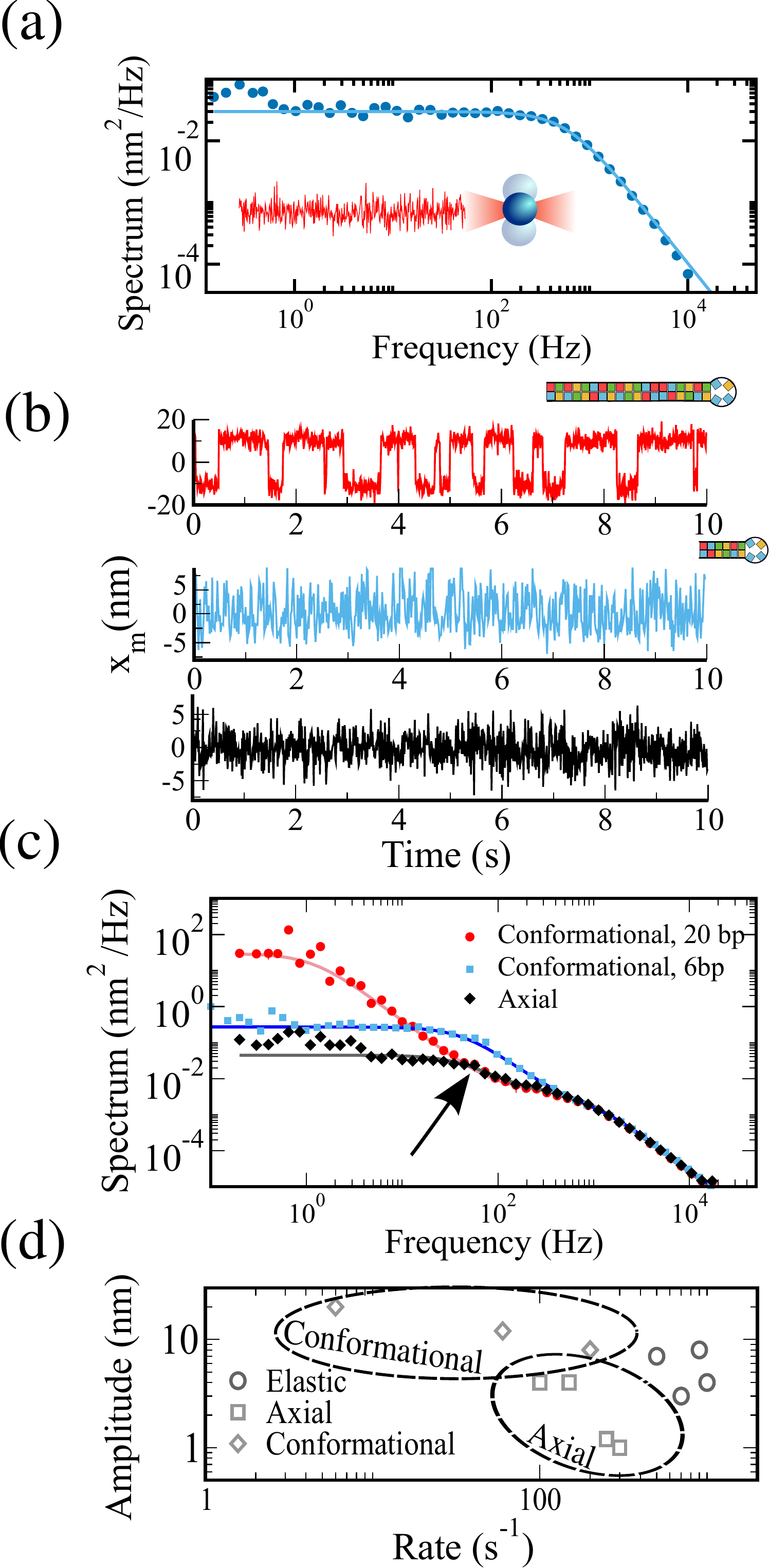}}
\caption{{\bf Spectral properties of OT force measurements.} a) Position fluctuation spectra of an optically trapped bead. The spectrum has a Lorentzian shape as expected for a Brownian particle in an harmonic potential well. b) Three
different traces illustrating noise measurements in OTs. Top: a 20bp hairpin under tension. Elastic and conformational fluctuations are clearly distinguishable. Center: a 6bp hairpin under tension. Elastic fluctuations mask conformational fluctuations. Bottom: elastic and axial fluctuations at high tension, where conformational dynamics is suppressed. All
extensions are relative and have zero mean. c) Spectra of the traces in panel b. Circles, squares, diamonds correspond to top, center, bottom. 
Continuous lines are double Lorentzian fits to the data. 
d) Amplitude and rates for the different noise components as measured in our dual-trap OT setup. Elastic fluctuations (circles) are well separated and clearly distinguishable. On the contrary conformational (diamonds) and axial (squares) fluctuations have similar rates.}
\end{figure}
Axial fluctuations are a manifestation of the three-dimensional nature of OT setups as opposed to one-dimensional idealizations often used in modeling. 
They can be identified decomposing the force signal in different orthogonal component as in \cite{ribezzi2012force}. Here we propose a different approach to identify 
axial fluctuations using only their projection on the force signal along the tether.
In Fig.~2d we show the typical amplitudes and timescales for different noise sources recorded in our OT setup (Fig. 1d). In the case of dual traps we will always be considering the spectrum of the differential signal \cite{PNAS.moffitt.2006,ribezzi2012force}. While the exact values for the amplitudes and timescales are  certainly  determined by the specific features of our setup and our tethers, the existence of three separate sectors for elastic, conformational and axial fluctuations is very likely a universal feature. In particular it is important to notice that, while the points corresponding to elastic fluctuations (circles) are well isolated in the diagram, conformational (diamonds) and axial (squares) fluctuations can have similar timescales. 
As a consequence, when observing the spectra in Fig.~2c (diamonds) the question arises whether the observed double Lorentzian behavior is due to axial fluctuations or to conformational dynamics. This is especially relevant when dealing with proteins which can have several intermediate states between the native and the coil state. A simple and general answer can be obtained by studying the force dependence of the corner frequency of the slow noise component 
(the fast noise is necessarily due to elastic fluctuations). In fact, the decay rate of axial fluctuations depends on force in a universal fashion, whose structure depends neither on the kind of setup  (single or dual trap) nor on the kind of tethered molecule:
\begin{equation}\label{only}
\omega_A(f)=\gamma_z^{-1}\left(k_z+\alpha\frac{f}{\ell}\right),
\end{equation}
where the parameters $\gamma_z$, $k_z$, $\alpha$ and $\ell$ depend on the setup and the molecular tether.   
Here $\gamma_z$ is the friction coefficient affecting axial fluctuations, $k_z$ is the axial stiffness of the optical trap,  $\ell$ is a characteristic length  and $\alpha$ 
is a numerical factor which takes the value 1 for single-trap and 2 for dual-trap setups. 
\begin{figure}[htp]
\centerline{\includegraphics[width=.70\columnwidth]{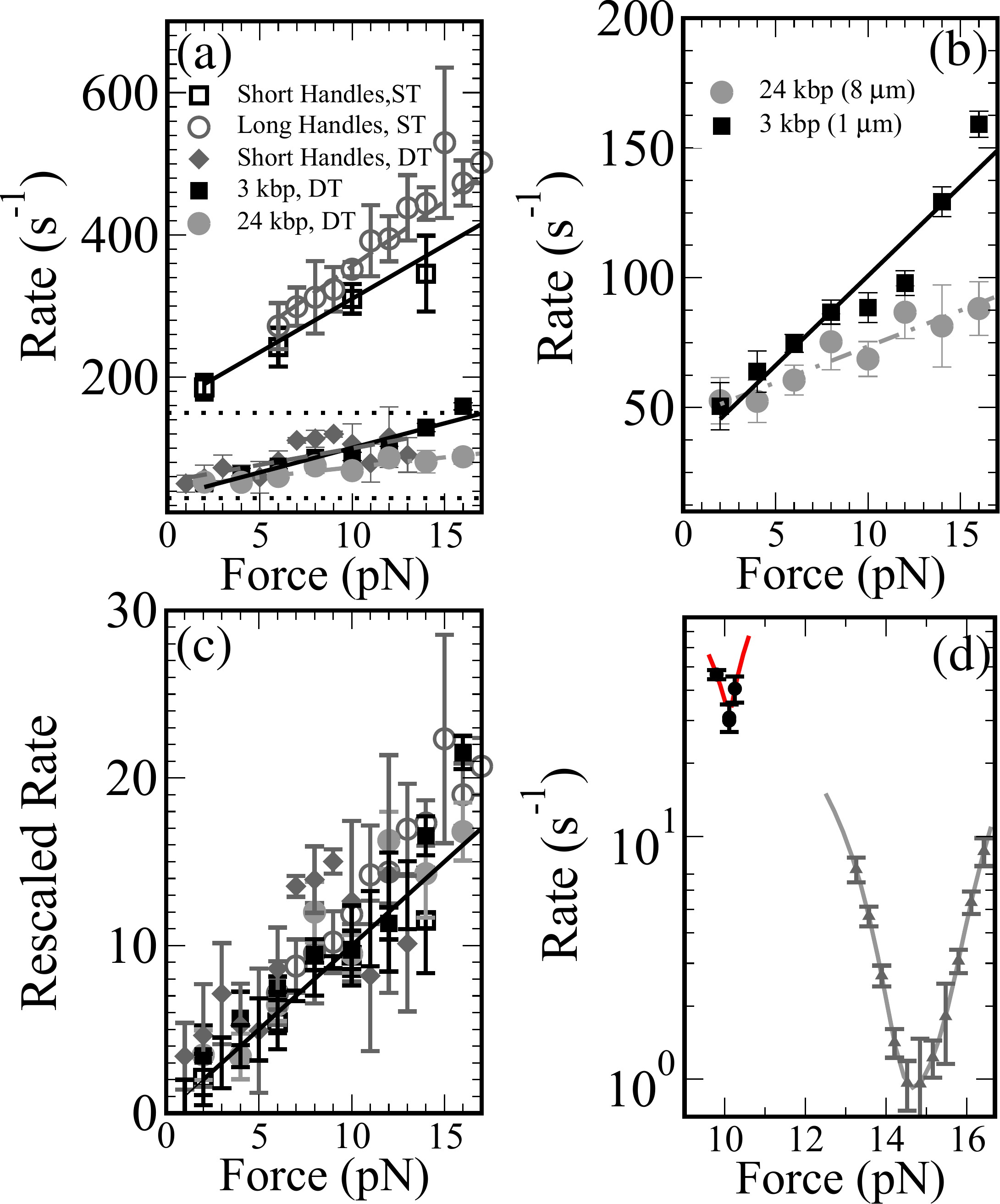}}
\caption{{\bf Decay rate of Axial Fluctuations.} (a) Corner frequency of axial fluctuations in single (ST) and dual (DT) OT setups for different tether lengths. 
Solid lines are fits of Eq.~\eqref{only} to the data. Fit results are reported in Table~\ref{tabulla}. (b) Corner frequency dependence on tether length. 
The frequency increases faster with force for shorter tethers  (smaller $\ell$). (c) Normalized rate $\tilde{\omega}_A$ for different values of $k_z$,$\ell$s and $\gamma_z$.
Rescaled frequencies fall on the master curve $\tilde{\omega}_A = f$. (d) Corner frequency of conformational fluctuations in a 6bp (circles) and a 20 bp hairpin (triangles) 
as a function of force. The characteristic chevron shape makes them clearly distinguishable from axial fluctuations.}
\end{figure}
In the case of dual trap setups $\ell$ is the length of the tether
plus twice the bead radius $\ell=x_m+2r_B$, while in single trap setups it is just the length of the tether plus the bead radius $\ell=x_m+r_B$ because the bead in the pipette cannot move. 
Equation ~\eqref{only} has been tested by studying the frequency of axial fluctuations as a function of force in both the single and the dual trap setups. In Fig.~3a we show the force dependence of the frequency of axial fluctuations measured on DNA hairpins with short and long handles in the single-trap mode and on different dsDNA 
tethers in the dual-trap mode. Tether lengths and bead radii are reported in Table 1, together with the result of fitting Eq.~\eqref{only} to the data. Imposing a value for $\ell$ it is possible to retrieve the axial trap stiffness $k_z$ and the viscosity parameter $\gamma_z$. The obtained values $\gamma_z$ and $k_z$ are compatible with measurements based on the spectrum of axial displacements of untethered beads (Beads only in Table~1). Note that in our setup the axial stiffness is ~10 times lower than the transverse stiffness (as reported in~\cite{ribezzi2013counter}) whereas the reduction is typically ~4-6 times in other setups \cite{RevSciInst.neuman.2004}. 
This is due to the fact that our setup uses laser beams underfilling the objectives as required for direct force measurement based on linear momentum conservation \cite{MEnzy.smith.2003}.
Although our single-trap is obtained using two counterpropagating laser beams Eq. \eqref{only} is also applicable to single-trap setups using a single laser beam.
In Fig.~3b we show how the force dependence of $\omega_A$ is affected by the tether length, $x_m$ ($x_m\simeq 8\mu$m for the 24kbp tether and $\simeq 1\mu$m for 3 kbp): shorter tethers (lower $x_m$) lead to a steeper increase of $\omega_A$ with force ($\ell=x_m+2r_B$).  
In Fig.~3c we show how the different datasets in Fig. 3a fall on the same master curve $\tilde{\omega}_A=f$ once they have been rescaled
for the different $\gamma_z,k_z,\ell$ values. The rescaling has been performed using the predicted value of $\ell$ and the values of $\gamma_z$ and $k_z$ obtained from the untethered bead measurements. Most importantly the force dependence of the decay rate of axial fluctuations is completely different from that of conformational fluctuations (Fig. 3d). The decay rate of two-state conformational fluctuations, $\omega_{CF}$ is given by
the sum of the folding and unfolding rates, $k_F,k_U:\,\omega_{CF}(f)=k_F(f)+k_U(f)$.
At low forces the sum is dominated by the folding rate, which decreases with force. Conversely at high forces this sum is dominated by the unfolding rate, which increases with force. 
As a consequence the corner frequency of conformational fluctuations depends on force in a non-monotonic way (Fig. 3d). The plot of measured rate vs force has a V-like shape.

In sum, we have discussed noise measurement in OTs and have highlighted axial fluctuations as an instrumental noise contribution that has been overlooked. The spectrum of axial fluctuations is such that they could be misinterpreted as conformational fluctuations in 
molecular systems. We provide the universal (independent of the setup) force dependence behavior of axial fluctuations. The strikingly different behavior of conformational and axial fluctuations under force (linear versus V-like) 
provides a general strategy to assess whether a Lorentzian shoulder corresponds to a genuine conformational transition or it is just instrumental noise.

F.R. is supported by an Instituci\'o Catalana de Recerca i Estudis Avançats Academia 2013 grant. The research leading to
these results has received funding from the European Union Seventh Framework Programme (FP7/2007-2013) under Grant 308850 INFERNOS
(information, fluctuations, and energy control in small systems).

\end{document}